\documentclass[final]{aipproc}

\layoutstyle{6s}

\begin{document}

\title{A New Astrophysical ``Triptych'': GRB030329/SN2003dh/URCA-2}

\author{M.G. Bernardini}{
  address={ICRA - International Center for Relativistic Astrophysics and Dipartimento di Fisica, Universit\`a di Roma ``La Sapienza'', Piazzale Aldo Moro 5, I-00185 Roma, Italy.}
}

\author{C.L. Bianco}{
  address={ICRA - International Center for Relativistic Astrophysics and Dipartimento di Fisica, Universit\`a di Roma ``La Sapienza'', Piazzale Aldo Moro 5, I-00185 Roma, Italy.}
}

\author{P. Chardonnet}{
  address={Universit\'e de Savoie, LAPTH - LAPP, BP 110, F-74941 Annecy-le-Vieux Cedex, France.}
}

\author{F. Fraschetti}{
  address={Universit\`a di Trento, Via Sommarive 14, I-38050 Povo (Trento), Italy.}
}

\author{R. Ruffini}{
  address={ICRA - International Center for Relativistic Astrophysics and Dipartimento di Fisica, Universit\`a di Roma ``La Sapienza'', Piazzale Aldo Moro 5, I-00185 Roma, Italy.}
}

\author{S.-S. Xue}{
  address={ICRA - International Center for Relativistic Astrophysics and Dipartimento di Fisica, Universit\`a di Roma ``La Sapienza'', Piazzale Aldo Moro 5, I-00185 Roma, Italy.}
}

\begin{abstract}
We analyze the data of the Gamma-Ray Burst/Supernova GRB030329/SN2003dh system obtained by HETE-2 (\citet{gcn1997}), R-XTE (\citet{gcn1996}), XMM (\citet{ta03}) and VLT (\citet{ha03}) within our theory (\citet{Brasile} and references therein) for GRB030329. By fitting the only three free parameters of the EMBH theory, we obtain the luminosity in fixed energy bands for the prompt emission and the afterglow (see Fig.~\ref{fig1}). Since the Gamma-Ray Burst (GRB) analysis is consistent with a spherically symmetric expansion, the energy of GRB030329 is $E = 2.1\times10^{52}$ erg, namely $\sim 2 \times 10^3$ times larger than the Supernova energy. We conclude that either the GRB is triggering an induced-supernova event or both the GRB and the Supernova are triggered by the same relativistic process. In no way the GRB can be originated from the supernova. We also evidence that the XMM observations (\citet{ta03}), much like in the system GRB980425/SN1998bw (\citet{cospar02,cospar02p}), are not part of the GRB afterglow, as interpreted in the literature (\citet{ta03}), but are associated to the Supernova phenomenon. A dedicated campaign of observations is needed to confirm the nature of this XMM source as a newly born neutron star cooling by generalized URCA processes.
\end{abstract}

\maketitle

A distinctive feature of our model, developed in the framework of the three interpretational paradigms (\citet{lett1,lett2,lett3}), has been the relation between the photon arrival time at the detector $t_a^d$ and the photon emission time $t$ (see \citet{Brasile,lett2,rbcfx02_letter}):
\begin{equation}
t_a^d = \left(1+z\right)\left(t - \frac{\int_0^tv\left(t'\right)dt'+r^\star}{c}\cos\vartheta + \frac{r^\star}{c}\right)\, ,
\label{tad}
\end{equation}
where $r(t)$, $v(t)$ and $\gamma(t)$ are the radial coordinate, the velocity and the Lorentz gamma factor of the expanding shell, $r^\star = r (t = 0)$, $\vartheta$ is the angle between the velocity of the emission point of the photon and the line of sight and $z$ is the cosmological redshift of the source.

In contrast with the relation between $t_a^d$ and $t$ used in the literature, which depends on an instantaneous value of the Lorentz $\gamma$ factor (see e.g. \citet{mr92}, Eq.(30) in \citet{p99}], Eq.(2) in \citet{vpkw00}, Eq.(2) in \citet{m02}), Eq.\eqref{tad} contains an integral which is a function of all previous values of the Lorentz gamma factor along the source world-line since the time $t = 0$. Therefore the knowledge of the Equations Of Motion (EOM) of the source is crucial to the evaluation of Eq.\eqref{tad}. In turns all the quantities which are computed using the EQuiTemporal Surfaces (EQTS, \citet{Brasile,rbcfx02_letter,eqts}) determined from Eq.\eqref{tad} become themselves very sensitive functions of the EOM. This includes the slope of the afterglow (\citet{Brasile}), which is essential in assessing the possible presence of beaming in the source (\citet{ta_prl}), the luminosity in fixed energy bands and the spectral analysis (\citet{Spectr1}).

\begin{figure}
\centering
\includegraphics[width=\hsize]{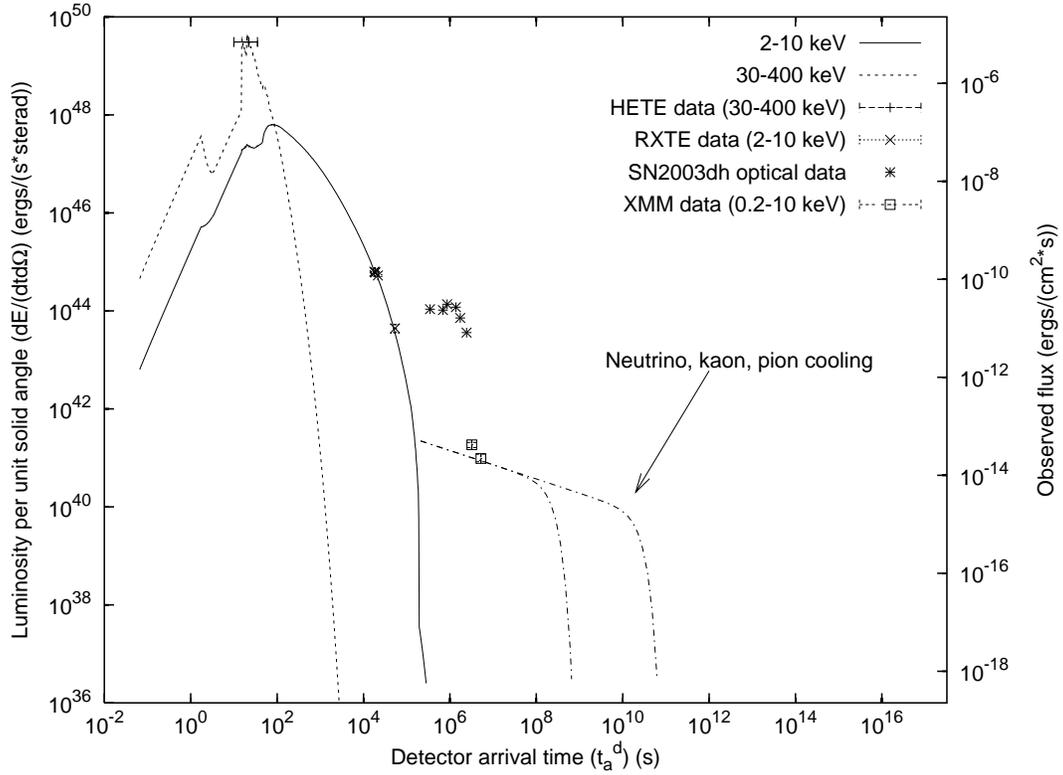}
\caption{The dotted line represents our theoretically predicted GRB030329 light curve in $\gamma$-rays (30-400 keV) with the horizontal bar corresponding to the mean peak flux from HETE-2 (\citet{gcn1997}). The solid line represents the corresponding one in X-rays (2-10 keV) with the experimental data obtained by R-XTE (\citet{gcn1996}). The remaining points refer respectively to the optical VLT data (\citet{ha03}) of SN2003bw and to the X-ray XMM data (\citet{ta03}) of URCA-2. The dash-dotted lines corresponds to cooling theoretical curves of young neutron stars by generalized URCA processes. It is interesting to compare and contrast these results with the ones for GRB980425/SN1998bw (see Fig. 3 in \citet{cospar02})}
\label{fig1}
\end{figure}

The determination of the EOM leads to a quite complex treatment, which starts from a very special set of initial conditions, proven to be unique. This treatment fits the observed luminosities with a large number of redundancy checks on the EOM (see Fig.~\ref{fig1}). It fits as well the time variability in the prompt radiation self-consistently with the determination of the EOM \citep{030329}.

We have adopted a spherically symmetric distribution for the GRB source and, as initial conditions at $t = 10^{-21}$ s, an $e^+$-$e^-$-photon neutral plasma lying between the radii $r_1 = 2.9\times 10^6$ cm and $r_2 = 9.0\times 10^7$ cm. The temperature of such a plasma is 2.1 MeV, the total energy $E_{tot} = 2.1\times 10^{52}$ erg and the total number of pairs $N_{e^+e^-} = 1.1\times 10^{57}$. These conditions have been derived evaluating the vacuum polarization processes (\citet{dr75}) occurring in the dyadosphere of an EMBH (\citet{rukyoto,prx98,crv02,rv02,rv03,rvx03}). The total energy $E_{tot}$ coincides with the dyadosphere energy $E_{dya}$ which is the first independent parameter of the EMBH theory. The optically thick electron-positron plasma created in the dyadosphere self-propels itself outward reaching ultrarelativistic velocities (\citet{rswx99}) and then interacts with the baryonic matter of the remnant of the progenitor star. The baryonic matter component $M_B$ is the second free parameter of the EMBH theory: $B = M_Bc^2/E_{dya} = 4.8 \times 10^{-3}$. The $e+$-$e^-$-photon-baryon plasma by further expansion becomes optically thin (\citet{rswx00}). As the transparency condition is reached, the Proper-GRB (P-GRB) is emitted with an extremely relativistic shell of Accelerated Baryonic Matter (the ABM pulse) with initial Lorentz gamma factor of $\gamma = 183.6$. It is this ABM pulse which produces the afterglow through its interaction with the ISM, whose average density is best fitted by $<n_{ism}> = 1$ particle/cm$^3$. In such a collision the ``fully radiative condition'' is implemented (for details see \citet{Brasile}): the internal energy $\Delta E_{int}$ which results is instantaneously radiated away.

We have recently assumed that the radiation emitted in the collision between the ABM pulse and the ISM has a thermal spectrum measured in the ABM pulse comoving frame (\citet{Spectr1}). In our approach the source luminosity is derived from an infinite set of foliations of events on the EQTS, each one characterized by a different thermal spectrum in the comoving frame boosted by a different relativistic transformation obtained from the EOM. The third free parameter of the EMBH theory describes this process of generating the thermal spectrum in the comoving frame. It is given by $1.1\times 10^{-7} < R = A_{eff}/A_{abm} < 5.0 \times 10^{-11}$, where $A_{abm}$ is the ABM pulse external surface area and $A_{eff}$ is the ABM pulse effective emitting area.

We can then obtain for the GRB030329 the luminosities in given energy bands, computed in the range 2-400 keV with very high accuracy. Fig.~\ref{fig1} shows the results for the luminosities in the 30-400 keV and 2-10 keV bands. Subsequently, the theoretically predicted GRB spectra have been evaluated at selected values of the arrival time \citep{030329}.

We can now compare these results with those for GRB980425/SN1998bw (\citet{cospar02}). We conclude that:

a) The intensity of the GRB versus the Supernova, comparable in the case of GRB980425, becomes $2\times 10^3$ times larger in the case of GRB030329. This crucial fact clearly indicates beyond any doubt the independence of the GRB phenomenon from the Supernova (\citet{lett3}). Moreover, the GRB is generally energetically dominant on the supernova; either the GRB is triggering an induced-supernova event or both the GRB and the Supernova are triggered by the same relativistic process. In no way the GRB can originate from the supernova.

b) In both systems the XMM observations point to the existence of an additional X-ray source, which we consider related to the Supernova phenomenon and not to the GRB. There is the distinct possibility that this source originates from the emission of a newly formed hot neutron star, cooling via generalized URCA processes (\citet{cospar02}). It has been recently proposed (\citet{r03mg10}) to indicate this new physical and astrophysical systems as URCA-1 for GRB980425/SN1998bw and URCA-2 for GRB030329/SN2003dh. A dedicated campaign of observations with XMM is urgently needed in order to explore this unprecedented ``triptych'' astrophysical systems, formed by a GRB, an induced-supernova and possibly a newly born pulsating hot neutron star. 

Details of this results are going to be published in \citep{030329}.

\bibliographystyle{aipproc}

\bibliography{bernardini_maria_grazia_0}

\end{document}